\documentclass[twocolumn,showpacs,preprintnumbers,prb,amsmath,amssymb]{revtex4}
\usepackage{graphicx}
\usepackage{amssymb}
\usepackage{bm}

\begin{document}

\title{Monte Carlo simulation with Tensor Network States}

\author{Ling Wang}
\affiliation{Faculty of physics, Boltzmanngasse 5, 1090 Vienna, Austria}

\author{Iztok Pi\v{z}orn}
\affiliation{Faculty of physics, Boltzmanngasse 5, 1090 Vienna, Austria}

\author{Frank Verstraete}
\affiliation{Faculty of physics, Boltzmanngasse 5, 1090 Vienna, Austria}

\date{\today}

\begin{abstract}
  We demonstrate that Monte Carlo sampling can be used to efficiently
  extract the expectation value of projected entangled pair states
  with a large virtual bond dimension. We use the simple update rule
  introduced by Xiang et al. in Phys.~Rev.~Lett {\bf 101}, 090603
  (2008) to obtain the tensors describing the ground state
  wavefunction of the Antiferromagnetic Heisenberg model and evaluate
  the finite size energy and staggered magnetization for square
  lattices with periodic boundary conditions of linear sizes up to
  $L=16$ and virtual bond dimensions up to $D=16$. The finite size
  magnetization errors are $0.003(2)$ and $0.013(2)$ at $D=16$ for a
  system of size $L=8,16$ respectively. Finite $D$ extrapolation
  provides exact finite size magnetization for $L=8$, and reduces the
  magnetization error to $0.005(3)$ for $L=16$, significantly
  improving the previous state of the art results.
\end{abstract}


\maketitle

The efficient simulation of strongly correlated quantum many body
systems presents one of the major open problems and challenges in
condensed matter physics. A major step forward was made by Steven
White ~\cite{whitedmrg} in the case of 1 dimensional quantum spin
chains by introducing the density matrix renormalization group (DMRG),
which soon became the method of choice for simulating 1 dimensional
manybody systems at zero temperature. By reformulating DMRG as a
variational method within the class of matrix product states (MPS)
\cite{ostlundandrommer,nishinovariational,VerstraetePorrasCirac}, it
has become clear how DMRG can be generalized to deal with systems in
two dimensions \cite{nishino-pre64-016705,frank0407066}; the quantum
states of the corresponding variational class are known as projected
entangled pair states (PEPS) and are part of the class called tensor
product states which also includes the multiscale entanglement
renormalization ansatz \cite{vidal-prl101-110501} and infinite PEPS
~\cite{jordan-prl101-250602}. More recently, it has also been
demonstrated how the PEPS class can take into account fermionic
anti-commutation relations
~\cite{kraus-pra81-052338,corboz-pra81-010303,corboz-prb80-165129,corboz-prb81-165104,barthel-pra80-042333,pizorn-prb-81-245110,gu-arxiv10042563}.
Numerical algorithms based on these ansatze, such as variational
minimization of the ground state energy and imaginary time evolution
are also developing fast
~\cite{frank0407066,frank-advphy57-143,jordan-prl101-250602,jiang-prl101-090603,gu-prb78-205116},
and a wide range of applications has been
studied~\cite{murg-prl95-057206,murg-pra75-033605,jordan-prb79-174515,murg-prb79-195119,xiang-prb81-174411,evenbly-arxiv09043383,bauer-jstatmech-09006,zhou-arxiv10013343,li-prb81-184427}.

The computational complexity of algorithms based on the PEPS ansatz
with virtual bond dimension $D$ scales as $D^{12}$ for the finite PEPS
algorithm with open boundary condition~\cite{frank-advphy57-143},
$\chi^{3}D^{4}$ for the infinite PEPS (iPEPS)
algorithm~\cite{jordan-prl101-250602}, $\chi^6$ for the tensor
entanglement renormalization (TERG) algorithm for square
lattices~\cite{gu-prb78-205116} and $\chi^{5}$ for honeycomb
lattices~\cite{gu-prb78-205116,jiang-prl101-090603}, where $\chi$ is
the number of Schmidt coefficients kept in the various
approximations. The large scaling power presents the main bottleneck
in scaling up the number of variational parameters, which is necessary
near second order phase transitions ~\cite{liu-prb82-060410}. The
common characteristics of all these algorithms is that the tensor
network is always contracted over the physical indices, which
effectively squares the computational cost of contracting the tensor
network as compared to a tensor network corresponding to a classical
spin system. As first shown in
\cite{schuch-prl100-040501,anders-prl99-220602} for the case of matrix
product states and string bond states, a square root speed up can be
obtained by using importance sampling over the physical indices. 
We will show how to adapt an importance sampling technique to PEPS.
The efficiency will depend on
the contraction algorithm chosen. In this paper we demonstrate it
using the TERG method.

The Antiferromagnetic Heisenberg model on a square lattice with length
$L$ has been well studied by stochastic series expansions
(SSE)~\cite{andersprb1997}, however it is notoriously hard for tensor
network wavefunctions to precisely capture the ground state order
parameter (staggered magnetization)
\cite{bauer-jstatmech-09006}. Various attempts have been made to
extract the right magnetization, {\it e.g.}  using iPEPS algorithm on
square lattice~\cite{bauer-jstatmech-09006} and the second
renormalization of tensor network state (SRG) on honeycomb
lattice~\cite{xiang-prb81-174411}. However, all those attempts
indicated that a tensor product state (TPS) with a finite $D$ has much
larger staggered magnetization in the thermodynamic limit. The main
reason for that is probably the fact that all TPS methods favour
states with a small amount of entanglement, and a larger local order
parameter indeed leads to states with a smaller amount of entanglement
due to the monogamy property of entanglement \cite{monogamy}. 


The simple update proposed in Ref.~\cite{jiang-prl101-090603} is an
extremely fast imaginary time evolution (projection) method, which
makes a simple estimation of the entanglement between the sub-system
and the environment and integrates it in the evolution step. The
evolution does not aim at the time dependent state at imaginary time
$\tau$; it aims at that, in the long run, the accumulative effect of
many non-sufficient improvements will eventually drive the system to
the ground state. Since there is no notion of the lattice size in this
update, one can claim the ground state obtained must be that of an
infinite lattice. Given a tensor product description of the
wavefunction with virtual bond dimension $D$ and its correlation
length $\xi(D)$, no local observable will have any notion the lattice
size if $L>\xi(D)$. We take the tensors obtained from the simple
update describing the ground state of Antiferromagnetic on an
\emph{infinite} lattice and evaluate the \emph{finite} size energy and
staggered magnetization with Monte Carlo (MC) sampling technique. We
show that the magnetization indeed reaches the correct value when
larger bond dimensions are used.

The paper is organized as following: in
Sec.~\ref{intro-terg} we give a brief introduction to the TERG algorithm,
in Sec.~\ref{sampling} we illustrate the sampling procedure using
the TERG contraction method, in Sec.~\ref{heisenberg} we apply the ground
state tensor obtained via the simple update (poorman's
update)~\cite{jiang-prl101-090603} to finite size lattices and
evaluate finite size expectation values via MC sampling, and finally a summary is given in Sec.~\ref{summary}.

\section{\label{intro-terg}Tensor entanglement renormalization algorithm}
\begin{figure}
\begin{center}
\includegraphics[width=8cm]{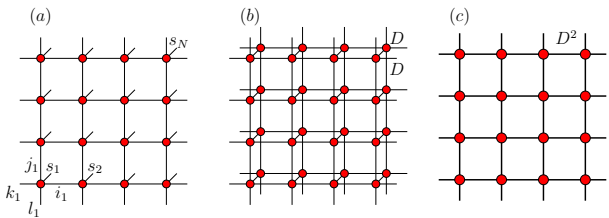}
\caption{(a) The tensor network wavefunction of a spin system on a
  square lattice, (b) The contraction of physical indices for
  calculating expectation values of a tensor network wavefunction,
  (c) this results in a tensor network of bond dimension $D^2$.}
\label{wavefunc1}
\end{center}
\end{figure}

The tensor network ansatz describes quantum many-body states in an
exponentially large Hilbert space in terms of local tensors
$\mathbf{T}$ describing local degrees of freedom. A graphical
representation of a tensor network state for a spin model on a square
lattice is presented in Fig.~\ref{wavefunc1}(a), for which the
wavefunction is written as
\begin{eqnarray}
\nonumber
|\psi\rangle
=\sum_{\{\sigma\}}{\mbox{tTr}\lbrace \mathbf{T}^{[1],s_1}\mathbf{T}^{[2],s_2}\cdots \mathbf{T}^{[N],s_N}\rbrace |s_1,s_2,\cdots,s_N\rangle},\\
\end{eqnarray}
where $\mathbf{T}^{[i],s_i}$ denotes the tensor of spin $s_i$ on site
$i$ and $|\sigma\rangle\equiv |s_1,s_2,\cdots,s_N\rangle$ represents
manybody spin configurations. The notation $\mbox{tTr}$ is used to
represent tensorial trace, generalization of the matrix trace to
tensor networks where tensors are traced over the virtual modes.

\begin{figure}
\begin{center}
\includegraphics[width=8cm]{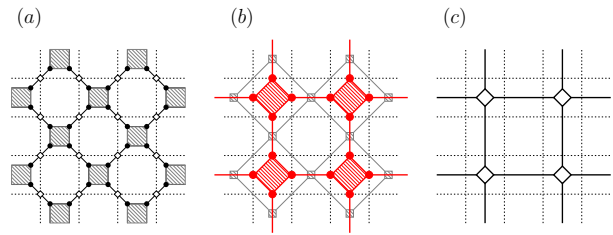}
\caption{(a) First decompose each 4-index tensor (in open diamonds on
  dash lines) into two 3-index tensors (in black dots), then contract
  every 4 tensors in the shaded area into a 4-index tensor (in open square
  in (b)). (b) Repeat the decomposition-contraction procedure on a
  reduced and rotated lattice (in gray solid line). (c) The total
  effect is that each 2 $\times$ 2 cluster on a fine lattice (in dash
  lines) is coarse grained into a super site (in open diamond) on a
  coarse grained lattice; note that the lattice orientation can be restored after every
  two iterations.}
\label{terg-steps}
\end{center}
\end{figure}

If one were to calculate an expectation value for a given observable
given a PEPS state, it would require contraction of a double layer
tensor network obtained by contracting over the physical bonds first
as depicted on Fig.~\ref{wavefunc1}(b). This results in a tensor
network with a squared bond dimension as shown on
Fig.~\ref{wavefunc1}(c).  The effort for contracting a tensor network
either as Fig.~\ref{wavefunc1}(a) for single layer or as
Fig.~\ref{wavefunc1}(c) for double layer grows exponentially with the
system size. One way of exact contraction is to successively
renormalize a $2\times 2$ block of sites into one super-site, as
pointed out in Ref.~\cite{ling-plq}. Without approximation, the
dimension of the virtual bond of a super-site in a double layer picture
will grow as $D^{4n_r}$, where $n_r$ is the depth of the
renormalization iteration (the linear size of the system is $L\sim
2^{n_r}$), thus approximate contraction becomes necessary. One of the
ways to contract the tensor network approximately is the tensor
renormalization method~\cite{levinandnave}, which was first proposed
to contract a classical tensor network. Later on this method was
generalized to deal with quantum
systems~\cite{gu-prb78-205116,jiang-prl101-090603}. The contraction
method on a square lattice can be described in
Fig.~\ref{terg-steps}. First each 4-index $\mathbf{T}$-tensor is
decomposed into two 3-index $\mathbf{S}$-tensors,
\begin{eqnarray}
\label{db}
T^{\textrm B}_{ijkl}=\sum_{\alpha}S^1_{ij\alpha}{S^{3}}_{kl\alpha},\\
\label{da}
T^{\textrm A}_{jkli}=\sum_{\alpha}S^2_{jk\alpha}{S^{4}}_{li\alpha},
\end{eqnarray}
where $\mathbf{T}^{\textrm{ B,A}}$ denote two ways of tensor
decomposition according to Eq.(\ref{db})(\ref{da}) respectively. In
the next step the four $\mathbf{S}$-tensors in the shaded area in
Fig.~\ref{terg-steps}(a) are contracted to form a coarse tensor on a
reduced and rotated lattice,
\begin{equation}
T^{\prime}_{\alpha\beta\gamma\delta}=\sum_{ijkl}S^2_{jk \alpha}S^3_{kl\beta}S^4_{li\gamma}S^1_{ij\delta}.
\end{equation}
This decomposition-contraction procedure can be applied once again on the
rotated lattice (Fig.~\ref{terg-steps}(b)) to obtain a coarse
lattice of half the length (Fig~\ref{terg-steps}(c)), and whose orientation
of the lattice is equal to the original one.

A singular value decomposition (SVD) is then done to decompose a
$\mathbf{T}$-tensor into two $\mathbf{S}$-tensors,
\begin{equation}
  T^{\textrm B}_{ijkl}=\sum_{\alpha=1}^{D^2}U_{ij\alpha}\Lambda_{\alpha}V_{kl\alpha}.
\label{svd_t}
\end{equation}
To prevent an exponential increase of the computational cost, one only keeps the largest
$D_{\textrm{cut}}$ (also referred to as $\chi$) singular values; this approximation
maximizes the 2-norm of vectorized $\mathbf{T}$ for a fixed $D_{\textrm{cut}}$,
\begin{equation}
\widetilde{T}^{\textrm B}_{ijkl}\approx\sum_{\alpha=1}^{D_{\textrm{cut}}}\bar{U}_{ij\alpha}\bar{\Lambda}_{\alpha}\bar{V}_{kl\alpha},
\end{equation}
where $\bar{\mathbf{M}}$ denotes taking the leading $D_{\textrm{cut}}$ columns of a matrix
$\mathbf{M}$ (or leading singular values of a diagonal matrix). A common
strategy is to absorb the diagonal matrix $\bar{\mathbf{\Lambda}}$ into
isometries $\bar{\mathbf{U}}$ and $\bar{\mathbf{V}}$ to obtain the $\mathbf{S}$-tensors,
$\mathbf{S}^1=\bar{\mathbf{U}}\sqrt{\bar{\mathbf{\Lambda}}}$ and
$\mathbf{S}^3=\bar{\mathbf{V}}\sqrt{\bar{\mathbf{\Lambda}}}$, the same applies to $\mathbf{S}^2,\mathbf{S}^4$.

\section{\label{sampling}Variational Quantum Monte Carlo sampling and update}

A variational quantum Monte Carlo (vQMC) method with tensor network
states can now be based on the TERG contraction method to calculate
the importance weight and the energy derivative. Following the
notation of Ref.~\cite{anders-prl99-220602}, we extract several key
equations regarding measuring and updating. For a chosen configuration
$|\sigma\rangle\equiv |s_1,s_2,\cdots, s_N\rangle$ we define a
coefficient $W(\sigma)=\langle\sigma|\psi\rangle$, which is calculated
by contracting a single layer tensor network as
\begin{equation}
W(\sigma)=\mbox{tTr}\lbrace \mathbf{T}^{[1],s_1}\mathbf{T}^{[2],s_2}\cdots \mathbf{T}^{[N],s_N}\rbrace.
\label{weights}
\end{equation}
The energy expectation value reads
\begin{equation}
\langle E\rangle=\frac{\sum_{\sigma}W^2(\sigma)E(\sigma)}{\sum_{\sigma}W^2(\sigma)},
\end{equation}
where
\begin{equation}
E(\sigma)=\sum_{\sigma^{\prime}}\frac{W(\sigma^{\prime})}{W(\sigma)}\langle \sigma^{\prime}|H|\sigma\rangle.
\end{equation}
The energy derivatives with respect to tensor elements $T^{\uparrow(\downarrow)}_{ijkl}$ are obtained via
\begin{equation}
\Big<\frac{\partial E}{\partial T^{\uparrow(\downarrow)}_{ijkl}}\Big>=2\langle\Delta^{\uparrow(\downarrow)}_{ijkl}(\sigma)E(\sigma)\rangle-2\langle\Delta^{\uparrow(\downarrow)}_{ijkl}(\sigma)\rangle\langle E(\sigma)\rangle,
\label{enrdrv}
\end{equation}
the $\langle\rangle$ denotes Monte Carlo average and
\begin{equation}
\Delta^{\uparrow(\downarrow)}_{ijkl}=\frac{1}{W(\sigma)}\frac{\partial W(\sigma)}{\partial T^{\uparrow(\downarrow)}_{ijkl}}.
\label{wsdrv}
\end{equation}
Let us define $\mathbf{B}(m)$ as the contraction of the tensor network for all sites except a site $m$:
\begin{equation}
\mathbf{B}(m)=\mbox{tTr}\lbrace\cdots \mathbf{T}^{[m-1],s_{m-1}}\mathbf{T}^{[m+1],s_{m+1}}\cdots\rbrace,
\label{envt}
\end{equation}
in terms of which we can express the derivative of the weight (\ref{weights}) with respect to
$T^{\uparrow(\downarrow)}_{ijkl}$ as
\begin{equation}
\frac{\partial W(\sigma)}{\partial T^{\uparrow(\downarrow)}_{ijkl}}=\sum_mB(m)_{ijkl}\delta_{s_m,{\uparrow(\downarrow)}},
\end{equation}
where we assume translation invariance symmetry, {\it i.e.}
$\mathbf{T}^{[i],s}=\mathbf{T}^{[j],s}$ ($s=\uparrow,\downarrow$) for
all sites $i,j$ in the lattice.

The program starts by randomly generating a spin configuration
$|\sigma\rangle$ satisfying $\sum_is_i=0$, {\it i.e.} we initialize
our state to live in total spin 0 sector. Given $|\sigma\rangle$, one
initializes and stores all the intermediate
$\mathbf{T}^{\textrm{q},\textrm{p}}$-tensor at the site $\textrm{q}$
of the $\textrm{p}^{\textrm{th}}$ coarse grained lattice. During the
contraction, we also calculate and store scalars
$f^{\textrm{q},\textrm{p}}\equiv
\textrm{max}\{|{T}^{\textrm{q},\textrm{p}}_{ijkl}|\}$ for all
$\mathbf{T}^{\textrm{q},\textrm{p}}$, then divide
$T^{\textrm{q},\textrm{p}}_{ijkl}$ by $f^{\textrm{q},\textrm{p}}$ to
avoid too large or too small singular values in the next
iteration. If we define the tensor trace of the final contraction step as
$g\equiv\mbox{tTr}\{\mathbf{T}^{[1],
  n_r}\mathbf{T}^{[2],n_r}\mathbf{T}^{[3],n_r}\mathbf{T}^{[4],n_r}\}$,
where $n_r$ is the number of iterations of a tensor network
contraction ($L=2^{\frac{n_r}{2}+1}$), then weight (\ref{weights}) can
be written as
\begin{equation}
W(\sigma)=g\prod_{\textrm{q},\textrm{p}}f^{\textrm{q},\textrm{p}}.
\label{weightform}
\end{equation}
Since we do not need to update the variational parameters
$T^{\uparrow(\downarrow)}_{ijkl}$ in this work, we will discuss how
to update a tensor network using energy derivatives with MC sampling
technique elsewhere~\cite{lingetal}.

While describing the sampling procedure, we take the nearest neighbor
Antiferromagnetic Heisenberg interaction as an example. Generalization
to other Hamiltonian is straight forward. Starting from site 1 of the
original tensor network, one looks for a pair of nearest neighbor
spins that align anti-parallel with each other and flip them. 

The trial configuration, which we denote as $|\sigma^{\prime}\rangle$, is
accepted with probability
\begin{equation}
P=\mbox{min}\left[ 1,\frac{W^2(\sigma^{\prime})}{W^2(\sigma)}\right],
\label{accrat}
\end{equation}
where the ratio is given by
\begin{equation}
\frac{W(\sigma^{\prime})}{W(\sigma)}=\frac{g^{\prime}}{g}\prod_{\textrm{q},\textrm{p}}\frac{f^{\prime \textrm{q},\textrm{p}}}{f^{\textrm{q},\textrm{p}}}.
\label{weightrat}
\end{equation}
To calculate the ratio (\ref{weightrat}), one needs to recompute some
$\mathbf{T}^{\prime \textrm{q},\textrm{p}}$ tensors together with the
corresponding scalars $f^{\prime \textrm{q},\textrm{p}}$ and
$g^{\prime}$, store them in separate arrays for later updates. If a
random number $r$ drawn from an uniform distribution on the interval
$[0,1)$ satisfies $r < P$, the trial state $|\sigma^{\prime}\rangle$
is accepted, in which case $\mathbf{T}^{\textrm{q},\textrm{p}}$,
$f^{\textrm{q},\textrm{p}}$ and $g$ are replaced by
$\mathbf{T}^{\prime \textrm{q},\textrm{p}}$,
$f^{\prime\textrm{q},\textrm{p}}$ and $g^{\prime}$, otherwise
$|\sigma^{\prime}\rangle$ is rejected and the original configuration
$|\sigma\rangle$ is kept. Moving through all the sites on the original
lattice, one attempts to flip all encountered anti-parallel pairs,
accepting or rejecting according to probability (\ref{accrat}). This
procedure is called a MC sweep. After each MC sweep, the energy and
other observables of interest are measured.  Flipping two neighboring
spins does not require recomputing many $\mathbf{T}^{\prime
  \textrm{q},\textrm{p}}$ tensors, which makes the contraction
fast. On the other hand, the update is local. To reduce the auto
correlation length, one needs to complete a MC sweep before making a
measurement, and the computational effort scales linearly with the
system size $N=L^2$.

\section{\label{heisenberg} The Antiferromagnetic Heisenberg model on square lattice}
We use the simple update method of Xiang et
al. \cite{jiang-prl101-090603} to obtain the converged wavefunction
with various virtual bond dimension ($D=3,4,\cdots,20$). The simple
update is an imaginary time evolution method to obtain the ground
state wavefunction of an infinite lattice. 

\begin{figure}
\begin{center}
\includegraphics[width=8cm]{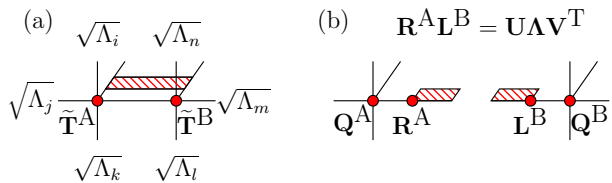}
\caption{(a) A evolution operator is applied to the nearest neighbor sites $\textrm{A,B}$ with weight $\sqrt{\Lambda_i}$ putting on the open bonds, (b) QR and LQ decompositions are done separately on $\widetilde{\mathbf{T}}^{\textrm{A}}$ and $\widetilde{\mathbf{T}}^{\textrm{B}}$  prior to the SVD step taken over the evolved bond between sites $\textrm{A,B}$. The computational cost is thus reduced from $D^9$ to $D^5$.}
\label{poorman}
\end{center}
\end{figure}

To implement the imaginary time evolution, we first make the Trotter
decomposition of the partition function $e^{-\beta H}=\lbrack e^{-\tau
  \sum_j H_j}\rbrack^{\textrm{M}}$ for $\tau=\beta/{\textrm{M}}$, then
we apply an evolution operator $e^{-\tau H_j}$ to the two nearest
neighbor sites $\textrm{A},\textrm{B}$ as in Fig.~\ref{poorman}(a). It
is crucial to put the weight $\sqrt{\Lambda_i}$ to the open auxiliary
modes to take into account the entanglement of the sub-system with the
environment. The weights $\sqrt{\Lambda_i}$ are the singular values
obtained in the previous evolution step on the corresponding
bond. According to Ref.~\cite{jiang-prl101-090603} an SVD is done to
the joint tensor
$\widetilde{\mathbf{T}}^{\textrm{A}}\widetilde{\mathbf{T}}^{\textrm{B}}$ to
introduce a cut over the enlarged bond, and only the leading $D$
singular values and the corresponding left and right eigenvectors are
kept as a projection to the sub-manifold of the Hilbert space where
the wavefunction manifests. Here we made a crucial modification that
drastically reduces the computational cost of decomposing the joint
matrix
$\widetilde{\mathbf{T}}^{\textrm{A}}\widetilde{\mathbf{T}}^{\textrm{B}}$ that
is of size $d\textrm{D}^{3}\times d\textrm{D}^3$. As illustrated in
Fig.~\ref{poorman}(b), we first make QR(LQ) factorization of tensor
$\widetilde{\mathbf{T}}^{\textrm{A(B)}}$ as
$\widetilde{\mathbf{T}}^{\textrm{A}}=\mathbf{Q}^{\textrm{A}}\mathbf{R}^{\textrm{A}}$
and
$\widetilde{\mathbf{T}}^{\textrm{B}}=\mathbf{L}^{\textrm{B}}\mathbf{Q}^{\textrm{B}}$,
where $\mathbf{R}^{\textrm{A}}$ and $\mathbf{L}^{\textrm{B}}$ are
right- and left-triangular matrices. Instead of a large tensor
$\widetilde{\mathbf{T}}^{\textrm{A}}\widetilde{\mathbf{T}}^{\textrm{B}}$, the
singular value decomposition is done on
$\mathbf{R}^{\textrm{A}}\mathbf{L}^{\textrm{B}}$, which is essentially
of size $d{\textrm{D}}\times d\textrm{D}$, as
$\mathbf{R}^{\textrm{A}}\mathbf{L}^{\textrm{B}}=\mathbf{U\Lambda
  V}^{\textrm{T}}$. The leading computational cost is now the QR(LQ)
decomposition that scales only to $D^5$. To obtain the evolved tensors
on sites $\textrm{A}$ and $\textrm{B}$, one has to remove the weight
$\sqrt{\Lambda_i}$ from the decomposed tensors as described in
Ref.~\cite{jiang-prl101-090603}.

In the thermodynamic limit, the ground state of the Antiferromagnetic
Heisenberg model spontaneously breaks the SU(2) symmetry, {\it i.e.}
the magnetization is locked in one direction. To achieve a significant
acceptance ratio for the Markov process with local spin flips, we
intentionally break the SU(2) symmetry into the XY plane. To do this,
we first attach a small ($h_a=0.001J$) staggered magnetic field in the
$x$ direction to the isotropic Heisenberg Hamiltonian,
\begin{equation}
H=J\sum_{\langle i,j\rangle}{\bm S}_i\cdot{\bm S}_j+h_a\sum_i(-1)^{i_x+i_y}S_i^x,\quad (J>0),
\end{equation}
here $i_x,i_y$ are $x,y$ coordinates of site $i$. We update the tensor
network wavefunction using this modified Hamiltonian until it
converges. Then we use the converged wavefunction to initialize a new
update without the field. The Trotter steps of the imaginary time
evolution is gradually reduced from $\tau=10^{-2}$ to $10^{-5}$. A
convergence is reached when
$\frac{\|\mathbf{T}^{\textrm{A(B)}}(\tau+100)-\mathbf{T}^{\textrm{A(B)}}(\tau)\|}{\|\mathbf{T}^{\textrm{A(B)}}(\tau)\|}<10^{-7}$,
where $\mathbf{T}^{\textrm{A,B}}(\tau)$ is the vectorized tensor at time slice
$\tau$, and is rescaled such that the largest magnitude of the tensor
elements is 1. We then take these converged tensors of various bond
dimension $D$ for the \emph{infinite} lattice to compute expectation values
of \emph{finite} lattices with periodic boundary condition using MC sampling
method.

\begin{figure}
\begin{center}
\includegraphics[width=7cm]{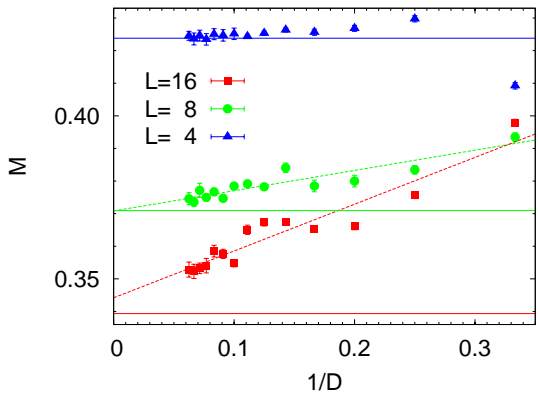}
\caption{Staggered magnetization as a function of $1/D$ for $L=4,8,16$. The solid lines are finite size expectation value from SSE and RVB projection methods. The dashed lines are linear fits for all bond dimensions $D$ for sizes $L=8,16$ respectively.}
\label{mag}
\end{center}
\end{figure}

One way to define the staggered magnetization is through the spin-spin
correlation at the longest distance~\cite{reger-prb37-5978}
\begin{equation}
M^2=\sum_{\alpha}C^{\alpha}(L/2,L/2),
\end{equation}
where $\alpha=x,y,z$, and
\begin{eqnarray}
\nonumber
C^{\alpha}(L/2,L/2)=\frac{1}{L^2}\sum_iS^{\alpha}(i_x,i_y)S^{\alpha}(i_x+\frac{L}{2},i_y+\frac{L}{2}).\\
\end{eqnarray}
In Fig.~\ref{mag}, we present the staggered magnetization as a
function of inverse virtual bond dimension $D$ for system sizes
$L=4,8,16$. The solid lines represents the magnetization results for
finite lattices obtained via SSE~\cite{andersprb1997} and resonating
valence bond (RVB) projection~\cite{andersprb82-024407} methods.  For
a small size $L=4$, large bond dimension $D\ge 8$ gives the exact
magnetization within the statistical error. For larger sizes $L=8,16$,
the magnetization error at $D=16$ is 0.003(2) 0.013(2) respectively.
Finite $D$ extrapolation gives the exact finite size magnetization for
$L=8$, and reduces the magnetization error to 0.005(3) for $L=16$.

\begin{figure}
\begin{center}
\includegraphics[width=7.cm]{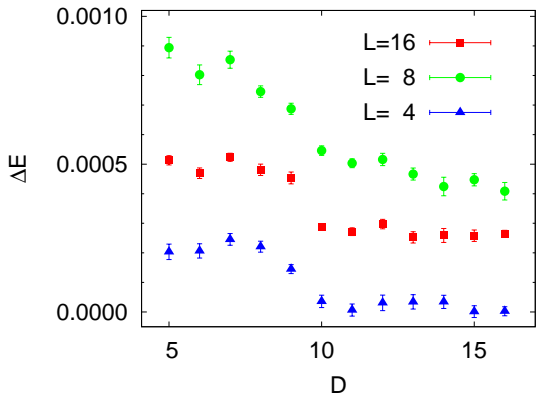}
\caption{Absolute error of the energy per bond as a function of the bond dimension $D$ for
  system sizes $L=4,8,16$ on a normal scale.}
\label{enrerr}
\end{center}
\end{figure}
In Fig.~\ref{enrerr} we present the absolute error of the finite size
energy divided by the number of bonds as a function of virtual bond
dimension $D$ for system sizes $L=4,8,16$. For all system sizes the
energy error drops significantly at $D=10$ and at $D\in [10:16]$
plateaus seem to set in.

\begin{figure}
\begin{center}
\includegraphics[width=7cm]{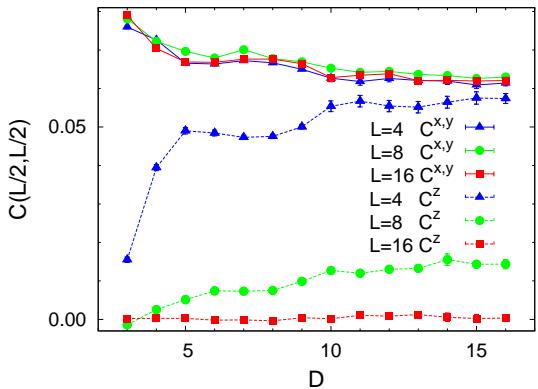}
\caption{Spin-spin correlation $C^{\alpha}(L/2,L/2)$ ($\alpha=x,y,z$)
  as a function of $D$ for system sizes $L=4,8,16$. Solid lines are
  the $x,y$ components, dash lines show the $z$ component. Note that
  $C^x=C^y$ for the $U(1)$ symmetry.}
\label{magccall}
\end{center}
\end{figure}
In Fig.~\ref{magccall} we show all three components of the spin-spin
correlation at the longest distance $C^{\alpha}(L/2,L/2)$,
$\alpha=x,y,z$. One can see that the $x,y$ components for different
system sizes almost fall on top of each other. The $x,y$ components
slightly drop at $D=5,10$ then followed by plateaus. The $z$
component, on the other hand, largely deviates from the $x$ and $y$
components. For $L=4$, the SU(2) symmetry is gradually restored with
increasing bond dimension $D$; for $L=8$, there is a partial growth of
$C^z(L/2,L/2)$ for increasing $D$; however for $L=16$, the $z$
component is zero for all available bond dimension
$D$. Asymptotically, as $D$ increase, one could expect
$C^{z}(L/2,L/2)$ grows to different values for different system sizes
$L$, and for sufficiently large system size $C^{z}(L/2,L/2)\to 0$ due
to automatic symmetry breaking.

\begin{figure}
\begin{center}
\includegraphics[width=7cm]{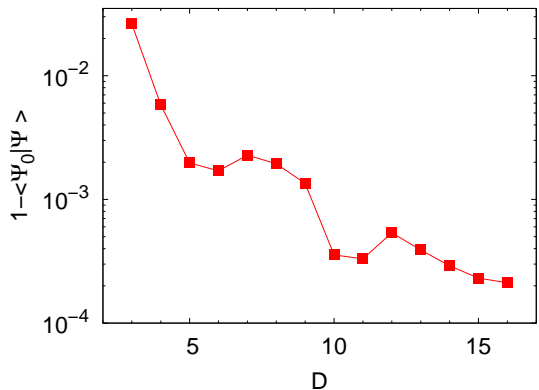}
\caption{Overlap of TPS $|\psi\rangle$ of various bond dimensions $D$
  with the exact ground state wavefunction $|\psi\rangle_0$ obtained
  by exact diagonalization for system size $L=4$. For $D=16$ the
  overlap is 0.99979.}
\label{overlap}
\end{center}
\end{figure}
In Fig.~\ref{overlap} we calculate the overlap of the TPS wavefunction
$|\psi\rangle$ for various bond dimensions $D$ with the exact ground
state $|\psi\rangle_0$ obtained by an exact diagonalization for a
$4\times 4$ system. For $D=16$ the overlap is 0.99979. 


For all the data presented, the maximum number of singular values
kept at each iteration step during the contraction is $D_{\textrm{cut}}=2D$ for
all bond dimensions $D$.

\section{\label{summary}Summary and discussion}
In this paper, we proposed a variational Quantum Monte Carlo (vQMC)
algorithm to evaluate a tensor network state of relative large bond
dimensions. We illustrated the Monte Carlo (MC) sampling procedure in
terms of the tensor entanglement renormalization group (TERG)
contraction algorithm. We applied this method to the well studied
Antiferromagnetic Heisenberg model on square lattice. Upon obtaining
the ground state wavefunction via imaginary time evolution with
essentially no notion of the lattice size, we evaluated the ground
state energy and the staggered magnetization for systems on
\emph{finite} square lattices using MC sampling method. Not
surprisingly, we found that the converged tensors obtained for the
infinite lattice give very accurate results also when used for
considered finite size lattices.
The wavefunction of a finite bond dimension $D$ thus can be used to
reliably extrapolate the expectation values in the thermodynamic limit
through finite $D$ scaling followed by finite size scaling. We have
shown that the tensor network ansatz based vQMC method is a promising
way to go to a very large bond dimension and thus allowing reliable
study of many interesting models.

We do not claim that the tensor product state (TPS) describing the
ground state wavefunction for an \emph{infinite} lattice obtained via
the simple update is the ultimate solution for a \emph{finite}
lattice. One still need further optimization for a finite lattice if
initializing from a TPS describing the infinite lattice. Many previous
studies of 1 dimensional systems had used the matrix product state
(MPS) obtained from an infinite chain algorithm~\cite{itebd} to
initialize the optimization for a finite chain and obtained remarkably
good results~\cite{shi1d,bogdan1d}. We will discuss how to update
tensors for a finite 2 dimensional system
elsewhere~\cite{lingetal}. Another advantage of the MC sampling method
is the possibility of incorporate lattice and spin symmetries into
the MC sampling scheme to improve accuracy, which has been
demonstrated in Ref.~\cite{sandvik2dmps} in the case of simulating 2
dimensional system via scale-renormalized MPSs. Since we obtained
tensor directly from the simple update, we did not employ symmetries
in the sampling procedure.

\acknowledgements We would like to thank A.~Sandvik, Z.-C.~Gu,
X.-G.~Wen, I.~Cirac, N.~Schuch, and H.-Q.~Zhou for useful
discussions. This project is supported by the EU Strep project
QUEVADIS, the ERC grant QUERG, and the FWF SFB grants FoQuS and
ViCoM. The computational results presented have been achieved in part
using the Vienna Scientific Cluster (VSC).

\end{document}